\documentclass[11pt]{article}

\usepackage[margin=1in]{geometry}
\usepackage{newtxtext}
\usepackage{microtype}
\usepackage{amsmath, amssymb, amsfonts}
\usepackage{graphicx}
\usepackage{booktabs}
\usepackage{multirow}
\usepackage{siunitx}
\usepackage{algorithm}
\usepackage{algpseudocode}
\usepackage{enumitem}
\usepackage{xcolor}
\usepackage{hyperref}
\usepackage[numbers,sort&compress]{natbib}
\usepackage{tikz}
\usetikzlibrary{arrows.meta,calc,positioning}

\usepackage{cleveref}
\usepackage[numbers,sort&compress]{natbib}
\usepackage{tikz}
\usepackage{tikz-cd}
\usepackage{microtype}
\usepackage{pgfplots}
\pgfplotsset{compat=1.18}
\usepackage{subcaption}
\usetikzlibrary{arrows.meta,positioning,calc,decorations.pathreplacing,patterns,shadings}

\definecolor{SolidGray}{RGB}{220,220,220}
\definecolor{AccentOrange}{RGB}{214,120,20}
\definecolor{AccentBlue}{RGB}{70,90,230}

\tikzset{
    >=Latex,
    panelbox/.style={
        fill=SolidGray,
        rounded corners=1.5mm,
        draw=none
    },
    poreboundary/.style={
        draw=black,
        line width=0.9pt,
        fill=solidgray
    },
    offsetcurve/.style={
        draw=gray!65,
        dashed,
        line width=0.8pt
    },
    throatfill/.style={
        fill=AccentOrange!25,
        draw=AccentOrange,
        line width=0.9pt
    },
    ann/.style={
        font=\small
    },
    smallann/.style={
        font=\footnotesize
    }
}

\hypersetup{
  colorlinks=true,
  linkcolor=red,
  citecolor=red,
  urlcolor=red
}

\newtheorem{remark}{Remark}

\newcommand{\placeholderfig}[2]{%
\begin{figure}[t]
\centering
\IfFileExists{#1}{%
  \includegraphics[width=0.78\linewidth]{#1}%
}{%
  \begin{tikzpicture}
    \draw[rounded corners, thick, gray] (0,0) rectangle (11,5.7);
    \node[align=center, text width=0.8\linewidth] at (5.5,2.85)
    {\Large Placeholder for figure\\[0.5em]\texttt{\detokenize{#1}}\\[0.5em]
    Generate this file from the Python simulation and place it in \texttt{figures/}.};
  \end{tikzpicture}%
}
\caption{#2}
\end{figure}
}

\hypersetup{
  colorlinks=true,
  linkcolor=blue,
  citecolor=blue,
  urlcolor=blue
}

\setlist[itemize]{leftmargin=1.2em}
\setlist[enumerate]{leftmargin=1.4em}

\definecolor{darkblue}{rgb}{0.0, 0.0, 0.8}
\definecolor{darkred}{rgb}{0.8, 0.0, 0.0}
\definecolor{darkgreen}{rgb}{0.0, 0.8, 0.0}

\definecolor{poreblue}{RGB}{112,198,242}
\definecolor{solidgray}{RGB}{170,170,170}
\definecolor{outred}{RGB}{220,70,70}
\definecolor{edgeblack}{RGB}{20,20,20}

\title{
Topological Signatures of Diffusive Release in Porous Media
}

\author{
Donghan Kim\thanks{
Department of Mathematical Sciences, KAIST, Daejeon, Republic of Korea.
Email: \texttt{patrick6231@kaist.ac.kr}
}
}

\date{}

\begin{document}
\maketitle

\begin{abstract}
We used persistent homology to quantify the multiscale topological and geometric
organization of porous media, including solid connectivity and the formation of
loop-like and cavity-like structures across spatial scales. Through statistical analysis, we show that these topological and geometric features are closely associated with diffusion-driven release behavior in porous media. In particular, even within each target-porosity level, samples with richer topological features tend to exhibit long-tailed release, indicating that release behavior depends not only on the amount of pore space but also on the multiscale organization of the solid phase. We further show that persistent homology-based features can classify release-curve regimes using a simple classification model. Notably, feature extraction is substantially faster than finite element
diffusion simulations. Together, these results suggest that persistent homology provides a lightweight, interpretable, and geometry-based descriptor for screening diffusive release behavior in porous media.
\end{abstract}

\section{Introduction}

Diffusive release in porous media is strongly influenced by pore-scale
geometry. Connectivity, tortuosity, constrictivity, accessible boundary area,
and the spatial arrangement of solid obstacles can all affect how solutes escape
from an initially loaded porous domain
\cite{crank1975mathematics,grathwohl1998diffusion,tartakovsky2019diffusion}.
This geometry--transport relationship is important in applications ranging from
subsurface contaminant transport and filtration to biomaterials and controlled
drug release~\cite{sahimi2011flow,blunt2013pore,yao1971water,karageorgiou2005porosity,
siepmann2012modeling,ahuja2009porous}. Mechanistic mathematical and numerical simulations are widely used to study
how porous geometry affects transport and release, ranging from pore-scale
image-based simulations of flow and transport to continuum models of
controlled drug release
\cite{blunt2013pore,arifin2006mathematical,fordversypt2013mathematical,
mcginty2017mathematical}.

A practical challenge is that direct numerical simulation of transport in
three-dimensional porous geometries can be computationally expensive, especially
when many candidate structures must be evaluated
\cite{blunt2013pore,wang2021ml}.
In screening or design settings, early identification of qualitative release
profiles can help prioritize candidate structures before more detailed
simulation or experimental evaluation
\cite{bannigan2023machine,protopapa2025machine,
robles2026predicting}. This motivates the search for
structural descriptors that are interpretable, geometry-based, and informative
about release curve behavior.

Persistent homology, one of the most widely used tools in topological data
analysis, provides a multiscale summary of topological features in data,
including connected components, loops, and cavities~\cite{carlsson2009topology,ghrist2008barcodes,carlsson2004persistence,
edelsbrunner2002topological,zomorodian2004computing}. In porous media,
persistent homology has been used to characterize connectivity, pore-size
structure, percolation length scales, permeability, elasticity, and fluid
trapping
\cite{robins2016percolating,ushizima2012augmented,jiang2018pore,
herring2019topological,moon2019statistical}. These studies suggest that
topological summaries can capture geometric information that is not fully
represented by simple scalar descriptors such as porosity.

\paragraph{Our Contribution}
In this work, we used persistent homology to quantify the multiscale topology and
geometry of the solid phase in porous media. A porosity-controlled statistical analysis shows that release regimes have
distinct topological signatures even after stratifying samples by target
porosity. In particular, more complex
solid-phase topology is associated with delayed or long-tail release, indicating
that release behavior is not determined by porosity alone but also by the
multiscale organization of the solid phase. We further show that these topological summaries can classify early-fast,
late-release, and long-tail regimes using a simple multinomial logistic
regression model, while timing comparisons demonstrate that the proposed
persistent homology-based method is substantially less computationally
expensive than finite element method (FEM)-based diffusion simulations. Together, these results suggest that
persistent homology can serve as a lightweight, interpretable, and
geometry-based descriptor for screening release behavior.

\section{Diffusion in Porous Media}
\label{sec:diffusion-porous-media}

We model a porous sample inside a cubic container $\Omega=[0,L]^3\subset \mathbb{R}^3$,
where $L>0$ is the side length. The solid phase is represented by a finite
collection of possibly overlapping spherical grains contained in the interior
of the cube. Let
$a_m\in\Omega$ and $\rho_m>0$ denote the center and radius of the $m$-th grain,
respectively, and assume that each closed ball is contained in $\operatorname{int}(\Omega)$.
The solid phase is defined by
\[
    S=\bigcup_{m=1}^M \overline{B}(a_m,\rho_m) ,
\]
and the pore space is its complement in the container, $P=\Omega\setminus S.$

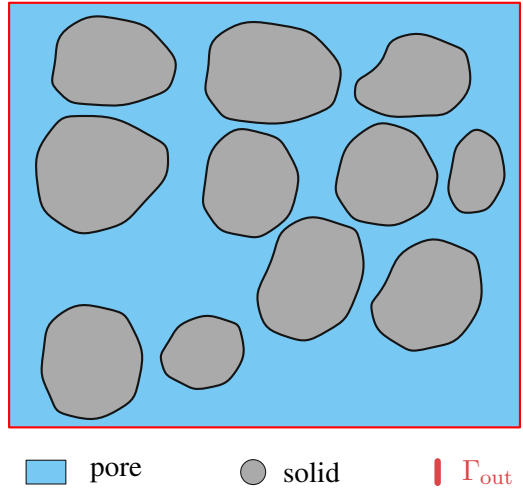
\begin{figure}[t]
\centering
\begin{tikzpicture}[scale=0.75, line cap=round, line join=round, >=Latex]


\tikzset{
    boxedge/.style={draw=edgeblack, line width=1.2pt},
    grain/.style={fill=solidgray, draw=edgeblack, line width=0.8pt},
}

\fill[poreblue!95] (0,0.5) rectangle (9,8);
\draw[red,thick] (0,0.5) rectangle (9,8);


\begin{scope}[shift={(0,0.2)}]
\filldraw[grain]
plot[smooth cycle, tension=0.95] coordinates
{(0.8,6.7) (1.1,7.4) (2.0,7.5) (2.7,7.1) (2.9,6.5) (2.4,6.1) (1.5,6.0) (0.9,6.2)};    
\end{scope}

\filldraw[grain]
plot[smooth cycle, tension=0.95] coordinates
{(3.5,6.8) (3.8,7.5) (4.8,7.6) (5.6,7.2) (5.8,6.5) (5.3,6.0) (4.2,5.9) (3.6,6.2)};

\begin{scope}[shift={(0.3,0.0)}]
\filldraw[grain]
plot[smooth cycle, tension=0.95] coordinates
{(6.2,6.9) (6.6,7.4) (7.4,7.3) (7.8,6.9) (7.6,6.2) (6.9,6.0) (6.1,6.1) (5.8,6.5)};
\end{scope}

\filldraw[grain]
plot[smooth cycle, tension=0.95] coordinates
{(0.5,4.8) (0.7,5.7) (1.4,6.0) (2.4,5.7) (2.8,5.2) (2.4,4.5) (1.7,4.0) (0.9,4.1)};

\begin{scope}[shift={(-0.35,-0.2)}]
\filldraw[grain]
plot[smooth cycle, tension=0.95] coordinates
{(3.8,5.0) (4.1,5.8) (4.8,5.9) (5.3,5.5) (5.4,4.8) (5.0,4.2) (4.4,4.1) (3.9,4.4)};
\end{scope}

\filldraw[grain]
plot[smooth cycle, tension=0.95] coordinates
{(5.8,5.0) (6.2,5.7) (6.9,5.8) (7.4,5.3) (7.5,4.7) (7.1,4.2) (6.4,4.1) (5.9,4.4)};

\filldraw[grain]
plot[smooth cycle, tension=0.95] coordinates
{(7.8,5.1) (8.0,5.6) (8.4,5.7) (8.7,5.2) (8.6,4.6) (8.2,4.3) (7.8,4.5)};


\filldraw[grain]
plot[smooth cycle, tension=0.95] coordinates
{(4.6,3.4) (5.0,4.1) (5.8,4.1) (6.2,3.7) (6.1,2.8) (5.6,2.2) (4.9,2.1) (4.4,2.6)};

\filldraw[grain]
plot[smooth cycle, tension=0.95] coordinates
{(6.6,3.0) (7.1,3.7) (7.9,3.7) (8.3,3.2) (8.1,2.3) (7.4,1.9) (6.8,2.0) (6.4,2.5)};

\filldraw[grain]
plot[smooth cycle, tension=0.95] coordinates
{(0.6,1.8) (0.9,2.5) (1.7,2.6) (2.2,2.2) (2.3,1.4) (1.9,0.8) (1.1,0.7) (0.7,1.1)};

\filldraw[grain]
plot[smooth cycle, tension=0.95] coordinates
{(2.8,2.0) (3.2,2.4) (3.8,2.4) (4.1,2.1) (4.0,1.5) (3.5,1.2) (3.0,1.3) (2.7,1.6)};




\begin{scope}[shift={(0.3,-0.5)}]
    \fill[poreblue!95] (0,0) rectangle (0.7,0.42);
    \draw[draw=edgeblack] (0,0) rectangle (0.7,0.42);
    \node[anchor=west] at (0.95,0.21) {pore};

    \fill[solidgray] (4,0.21) circle (0.22);
    \draw[draw=edgeblack] (4,0.21) circle (0.22);
    \node[anchor=west] at (4.35,0.21) {solid};

    \draw[line width=2.4pt, color=outred] (7.25,0.02) -- (7.25,0.40);
    \node[anchor=west, text=outred] at (7.5,0.21) {\(\Gamma_{\mathrm{out}}\)};
\end{scope}

\end{tikzpicture}
\caption{
A schematic porous medium in a square cross-section of the sample domain
$\Omega$. The gray inclusions represent solid grains $S$, shown as separated
for visual clarity. The blue region is the pore space $P=\Omega\setminus S$.
The red segment on the boundary indicates an absorbing outlet $\Gamma_{\mathrm{out}}$.
}
\label{fig:porous_medium_schematic}
\end{figure}

Tracer release is modeled by diffusion in the pore space. We prescribe an
absorbing outlet boundary $\Gamma_{\mathrm{out}}\subset\partial\Omega$, which
represents contact with a perfect exterior sink. In the fully exposed case one
may take $\Gamma_{\mathrm{out}}=\partial\Omega$, while partially coated samples
can be modeled by choosing $\Gamma_{\mathrm{out}}$ as a proper subset of
$\partial\Omega$. The remaining boundary is treated as reflecting; $\Gamma_{\mathrm{wall}}
    =
    \partial P\setminus \Gamma_{\mathrm{out}}.$
Thus $\Gamma_{\mathrm{wall}}$ includes the solid--pore interface and any
non-absorbing part of the exterior boundary (see \Cref{fig:porous_medium_schematic}).

Let $c(x,t)$ be the tracer concentration in $P$. We consider the mixed
Dirichlet--Neumann diffusion problem
\[
\begin{cases}
\displaystyle
\frac{\partial c}{\partial t}=D\Delta c,
& x\in P,\ t>0, \\[0.5em]
c(x,t)=0,
& x\in \Gamma_{\mathrm{out}},\ t>0, \\[0.5em]
\displaystyle
\frac{\partial c}{\partial \nu}(x,t)=0,
& x\in \Gamma_{\mathrm{wall}},\ t>0, \\[0.5em]
c(x,0)=1,
& x\in P.
\end{cases}
\]
Here $D>0$ is the diffusion coefficient and $\nu$ is the outward unit normal on
$\Gamma_{\mathrm{wall}}$. The Dirichlet condition represents irreversible tracer loss
through the outlet boundary, whereas the Neumann condition represents
impermeable solid walls and coated exterior regions.

The cumulative release curve is defined by
\[
    Q(t)
    =
    1-\frac{\int_P c(x,t)\,dx}{\int_P c(x,0)\,dx}.
\]
Since the initial concentration is uniform, this becomes
\[
    Q(t)
    =
    1-\frac{1}{|P|}\int_P c(x,t)\,dx.
\]
Therefore, $Q(t)$ measures the fraction of tracer that has escaped from the
pore space by time $t$.

\section{Methods}

\subsection{Finite Element Approximation of the Release Curve}
\label{subsec:fem-release}

The release curve $Q(t)$ is obtained by numerically solving the mixed
Dirichlet--Neumann diffusion problem on a voxelized approximation of the pore
space. Although the underlying solid phase is defined by spherical obstacles,
the computational pore domain is represented on a regular voxel grid. Each pore
voxel is subdivided into six tetrahedral elements, producing a voxel-derived
tetrahedral mesh on which a conforming finite element method is applied.

This construction should be distinguished from solving the diffusion equation
directly by a finite-difference scheme on the voxel grid. The voxel grid is used
to define the computational geometry, whereas the diffusion equation is solved
using linear finite elements on the resulting tetrahedral mesh. This approach
allows us to retain a finite element variational formulation while using a
simple and reproducible discretization of the pore geometry.

Let $P_h$ denote the voxelized pore domain and let
$\mathcal{T}_h$ be the tetrahedral mesh obtained by subdividing each pore voxel
into six tetrahedra. The absorbing part of the exterior boundary is denoted by
$\Gamma_{\mathrm{out},h}$, and the remaining boundary, including the voxelized
solid--pore interface, is denoted by $\Gamma_{\mathrm{wall},h}$. We impose a
homogeneous Dirichlet condition on $\Gamma_{\mathrm{out},h}$ and a homogeneous
Neumann no-flux condition on $\Gamma_{\mathrm{wall},h}$.

The finite element approximation is based on the weak formulation of the
diffusion equation. Define
\[
    V_h
    =
    \{v_h \in C^0(P_h): v_h|_K \in \mathbb{P}_1(K)
    \text{ for all } K\in \mathcal{T}_h,\;
    v_h=0 \text{ on } \Gamma_{\mathrm{out},h}\}.
\]
Here $\mathbb{P}_1(K)$ denotes the space of linear polynomials on a tetrahedral
element $K$. The homogeneous Dirichlet condition on the absorbing boundary is
imposed strongly through the definition of $V_h$, while the homogeneous
Neumann condition on the voxelized solid--pore interface is incorporated
naturally in the variational formulation.

For a time step $\Delta t>0$, we use backward Euler time discretization. Given
$c_h^{n-1}\in V_h$, we compute $c_h^n\in V_h$ by
\[
    \int_{P_h}
    \frac{c_h^n-c_h^{n-1}}{\Delta t} v_h\,dx
    +
    D
    \int_{P_h}
    \nabla c_h^n \cdot \nabla v_h\,dx
    =
    0
    \qquad
    \text{for all } v_h\in V_h .
\]
Equivalently, in matrix form this gives
\[
    (M+\Delta t\,D K)\mathbf{c}^n
    =
    M\mathbf{c}^{n-1},
\]
where $M$ and $K$ are the finite element mass and stiffness matrices assembled
on the tetrahedral mesh $\mathcal{T}_h$. The initial condition is obtained by
interpolating the uniform concentration $c(x,0)=1$ onto the finite element
space, with the prescribed zero values on the absorbing boundary nodes.

The numerical cumulative release curve is then defined by
\[
    Q_h(t_n)
    =
    1-
    \frac{\int_{P_h} c_h^n(x)\,dx}
         {\int_{P_h} c_h^0(x)\,dx},
    \qquad
    t_n=n\Delta t .
\]
Thus $Q_h(t_n)$ approximates the fraction of tracer that has escaped through
the absorbing exterior boundary by time $t_n$.

As the voxel resolution is refined, the voxelized
geometry provides a progressively finer approximation of the underlying porous
medium. The backward Euler finite element scheme is unconditionally stable for
linear diffusion problems, and standard convergence theory for conforming
Galerkin finite element methods applies to the corresponding discretized
domains under the usual mesh-regularity and solution-regularity assumptions
\cite{ciarlet2002finite,brenner2008mathematical,quarteroni2008numerical,thomee2006galerkin}.

\subsection{Persistent Homology}

Persistent homology provides a way to measure how long topological features,
such as connected components, loops, and voids, remain visible as the data are
viewed at progressively larger scales~\cite{carlsson2004persistence,
edelsbrunner2002topological,zomorodian2004computing}; see, e.g.,
\cite{edelsbrunner2010computational} for detailed background. In this study, we
use persistent homology to describe the multiscale topology of porous-media
geometry. The main idea is to gradually offset the solid phase and record how
its topological features change during this process.

Let $\Omega$ denote the image or computational domain, and let $S\subset\Omega$
be the solid phase. To describe the solid geometry at multiple spatial scales,
we consider the filtration $\{\mathcal{F}_r\}_{r\geq 0}$, where
\[
    \mathcal{F}_r
    =
    \{x\in\Omega : \operatorname{dist}(x,S)\le r\}.
\]
In other words, $\mathcal{F}_0=S$, and increasing $r$ gradually expands the
solid phase into the surrounding pore space. This allows us to observe how
separate solid components become connected, how loop-like structures appear,
and how enclosed regions are formed or disappear as the scale increases.

We quantify these changes using the homology groups $H_q(\mathcal{F}_r)$.
Here, $H_0$ describes connected solid components, $H_1$ describes loop-like
structures formed by the expanding solid phase, and $H_2$ describes enclosed
void-like regions surrounded by solid structures. In the context of porous
media, these features provide information about the spatial organization of
solid obstacles and the pore regions between them.

Each topological feature is assigned a birth $b_i$ and a death $d_i$. The birth scale records the value of $r$ at which the feature first
appears, while the death scale records the value of $r$ at which it disappears,
for example because it merges with another feature or is filled in. The
resulting birth--death pairs for the $q$th homology group form the persistence
diagram
\[
    \mathrm{Dgm}_q
    =
    \{(b_i,d_i)\}_{i=1}^{n_q}.
\]
Roughly speaking, the lifetimes $d_i-b_i$ in a persistence diagram provide scale
information about the corresponding topological features. For example, long
lifetimes in the $H_1$ or $H_2$ persistence diagrams indicate geometric or
topological structures that persist across a wide range of offset scales,
whereas short lifetimes correspond to small-scale or rapidly disappearing
features. These persistence diagrams therefore summarize multiscale information
about solid connectivity, pore bottlenecks, and chamber-like regions that may
influence the diffusion release curve $Q(t)$.

\begin{remark}[Computation by weighted alpha complexes]
The solid-offset filtration is the geometric object used throughout the paper.
At scale $\alpha$, the offset solid phase is
\[
    S_\alpha
    =
    \{x\in\Omega:\mathrm{dist}(x,S)\le \alpha\}
    =
    \bigcup_{m=1}^{M}\overline{B}(a_m,\rho_m+\alpha).
\]
Equivalently, this is the union of the balls
$\overline{B}(a_m,\rho_m+\alpha)$.

For each $\alpha$, let $\check{C}_\alpha$ denote the \v{C}ech complex of the
expanded balls $\{\overline{B}(a_m,\rho_m+\alpha)\}_{m=1}^M$~\cite{bell2019weighted}. Since finite
intersections of Euclidean balls are either empty or contractible, the nerve
theorem~\cite{borsuk1948imbedding} implies that $\check{C}_\alpha$ is homotopy
equivalent to $S_\alpha$. Moreover, these equivalences are compatible with the
inclusions as $\alpha$ increases, and hence the \v{C}ech filtration computes the
persistent homology of the solid-offset filtration.

For computation, we use the corresponding weighted alpha-complex filtration.
This filtration yields the same persistence diagrams as the weighted \v{C}ech
filtration of the expanded grains, while allowing the diagrams to be computed
more efficiently. Therefore, the persistence diagrams computed from the weighted
alpha filtration represent the persistent homology of the solid-offset
filtration.
\end{remark}

\begin{figure}
\centering

\begin{tikzpicture}[
    scale=1.2
]

\begin{scope}[shift={(0,0)}]
    \fill[
        poreblue,
        rounded corners=5pt
    ] (0,0) rectangle (2.5,1.75);
    \coordinate (P1) at (0.7,0.9);
    \coordinate (P2) at (1.25,1.2);
    \coordinate (P3) at (1.25,0.6);
    \coordinate (P4) at (1.9,1.2);
    \coordinate (P5) at (1.9,0.6);

    \fill[solidgray] (P1) circle (0.05);
    \fill[solidgray] (P2) circle (0.15);
    \fill[solidgray] (P3) circle (0.15);
    \fill[solidgray] (P4) circle (0.15);
    \fill[solidgray] (P5) circle (0.15);
    \draw (P1) circle (0.05);
    \draw (P2) circle (0.15);
    \draw (P3) circle (0.15);
    \draw (P4) circle (0.15);
    \draw (P5) circle (0.15);

    \node[below] at (1.25,0) {$r=r_1$};
\end{scope}

\begin{scope}[shift={(3,0)}]
    \fill[
        poreblue,
        rounded corners=5pt
    ] (0,0) rectangle (2.5,1.75);
    \coordinate (P1) at (0.7,0.9);
    \coordinate (P2) at (1.25,1.2);
    \coordinate (P3) at (1.25,0.6);
    \coordinate (P4) at (1.9,1.2);
    \coordinate (P5) at (1.9,0.6);

    \fill[solidgray] (P1) circle (0.15);
    \fill[solidgray] (P2) circle (0.25);
    \fill[solidgray] (P3) circle (0.25);
    \fill[solidgray] (P4) circle (0.25);
    \fill[solidgray] (P5) circle (0.25);
    \draw (P1) circle (0.15);
    \draw (P2) circle (0.25);
    \draw (P3) circle (0.25);
    \draw (P4) circle (0.25);
    \draw (P5) circle (0.25);
    \node[below] at (1.25,0) {$r=r_2$};
    \draw[
        ->,
        very thick,
        black,
        >=Stealth
    ] (-0.4,0.875) -- (-0.1,0.875);
\end{scope}

\begin{scope}[shift={(6,0)}]
    \fill[
        poreblue,
        rounded corners=5pt
    ] (0,0) rectangle (2.5,1.75);
    
    \coordinate (P1) at (0.7,0.9);
    \coordinate (P2) at (1.25,1.2);
    \coordinate (P3) at (1.25,0.6);
    \coordinate (P4) at (1.9,1.2);
    \coordinate (P5) at (1.9,0.6);

    \fill[solidgray] (P1) circle (0.2);
    \fill[solidgray] (P2) circle (0.35);
    \fill[solidgray] (P3) circle (0.35);
    \fill[solidgray] (P4) circle (0.35);
    \fill[solidgray] (P5) circle (0.35);
    \draw (P1) circle (0.2);
    \draw (P2) circle (0.35);
    \draw (P3) circle (0.35);
    \draw (P4) circle (0.35);
    \draw (P5) circle (0.35);

    \draw[red,thick] (P2)--(P3)--(P5)--(P4)--(P2);
    
    \node[below] at (1.25,0) {$r=r_3$};
    \draw[
        ->,
        very thick,
        black,
        >=Stealth
    ] (-0.4,0.875) -- (-0.1,0.875);
\end{scope}
\begin{scope}[shift={(1.5,-2.2)}]
    \fill[
        poreblue,
        rounded corners=5pt
    ] (0,0) rectangle (2.5,1.75);
    
    \coordinate (P1) at (0.7,0.9);
    \coordinate (P2) at (1.25,1.2);
    \coordinate (P3) at (1.25,0.6);
    \coordinate (P4) at (1.9,1.2);
    \coordinate (P5) at (1.9,0.6);

    \fill[solidgray] (P1) circle (0.25);
    \fill[solidgray] (P2) circle (0.4);
    \fill[solidgray] (P3) circle (0.4);
    \fill[solidgray] (P4) circle (0.4);
    \fill[solidgray] (P5) circle (0.4);
    \draw (P1) circle (0.25);
    \draw (P2) circle (0.4);
    \draw (P3) circle (0.4);
    \draw (P4) circle (0.4);
    \draw (P5) circle (0.4);

    \draw[red,thick] (P2)--(P3)--(P5)--(P4)--(P2);
    \draw[green,thick] (P2)--(P3)--(P1)--(P2);
    
    \node[below] at (1.25,0) {$r=r_4$};
    \draw[
        ->,
        very thick,
        black,
        >=Stealth
    ] (-0.4,0.875) -- (-0.1,0.875);
\end{scope}

\begin{scope}[shift={(4.5,-2.2)}]
    \fill[
        poreblue,
        rounded corners=5pt
    ] (0,0) rectangle (2.5,1.75);
    \coordinate (P1) at (0.7,0.9);
    \coordinate (P2) at (1.25,1.2);
    \coordinate (P3) at (1.25,0.6);
    \coordinate (P4) at (1.9,1.2);
    \coordinate (P5) at (1.9,0.6);

    \fill[solidgray] (P1) circle (0.3);
    \fill[solidgray] (P2) circle (0.45);
    \fill[solidgray] (P3) circle (0.45);
    \fill[solidgray] (P4) circle (0.45);
    \fill[solidgray] (P5) circle (0.45);
    \draw (P1) circle (0.3);
    \draw (P2) circle (0.45);
    \draw (P3) circle (0.45);
    \draw (P4) circle (0.45);
    \draw (P5) circle (0.45);

    \filldraw[red!40] (P2)--(P3)--(P5)--(P4)--(P2);
    \filldraw[green!40] (P2)--(P3)--(P1)--(P2);
    
    \node[below] at (1.25,0) {$r=r_5$};
    \draw[
        ->,
        very thick,
        black,
        >=Stealth
    ] (-0.4,0.875) -- (-0.1,0.875);

\end{scope}

\begin{scope}[shift={(10,-2)}, scale=1.2]
    \draw[->, thick] (0,0) -- (3.2,0) node[below left] {Birth};
    \draw[->, thick] (0,0) -- (0,2.65) node[above, rotate=90, anchor=south] {Death};

    \draw[thick] (0,0) -- (2.75,2.35);

    \draw[dashed, black!60] (0.5,0) -- (0.5,2.25);
    \fill[red] (0.5,2.25) circle (0.07);

    \draw[dashed, black!60] (1,0) -- (1,2.25);
    \fill[green] (1,2.25) circle (0.07);



    \node[below] at (0.5,0) {$r_3$};
    \node[below] at (1,0) {$r_4$};

    \draw[thick] (-0.04,0) -- (0.04,0);
    \draw[thick] (0,-0.04) -- (0,0.04);

\end{scope}

\end{tikzpicture}
\caption{New rings appear at $r_i$ $(i=3,4)$, and the filled rings indicate that these features disappear.}
\label{fig:persistent_homology}

\end{figure}
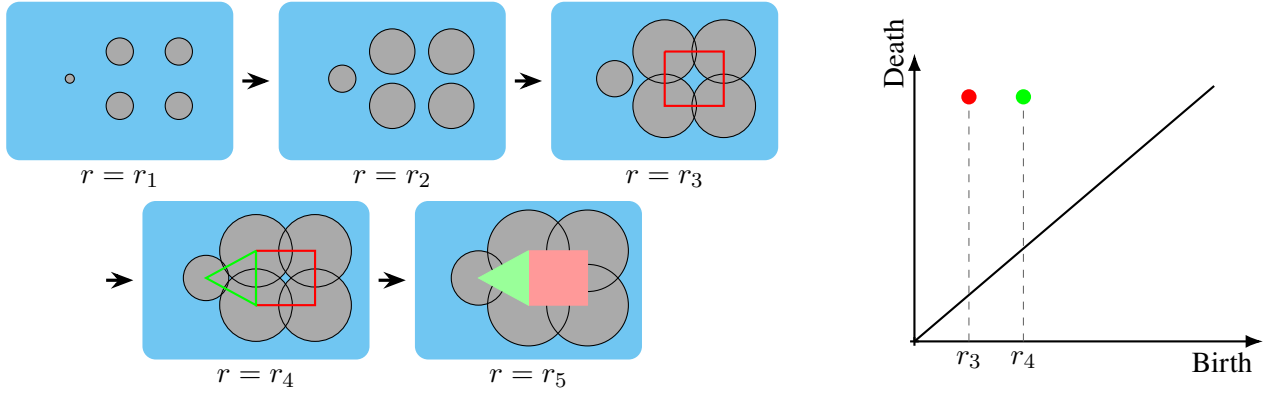


\section{Experimental Setting and Results}
\label{sec:experiments-results}

\subsection{Experimental Setting}
\label{subsec:experimental-setting}

We generate synthetic porous-media samples using the notation introduced in
Section~\ref{sec:diffusion-porous-media}. Throughout the experiments, the
computational domain is the unit cube $\Omega=[0,1]^3$, and the solid phase is
generated by placing spherical obstacles in the interior of $\Omega$. Sphere
overlaps are allowed, and each sample is generated to approximately match a
prescribed target porosity.

To obtain geometrically diverse samples, we use six structural classes:
random monodisperse, random polydisperse, channel-like, layered,
bottleneck-like, and clustered configurations. These classes are summarized in
Table~\ref{tab:structure-types}. Within each grid--porosity setting, the six
classes are sampled approximately uniformly.

\begin{table}[t]
\centering
\caption{Structural classes used to generate synthetic porous-media samples.}
\label{tab:structure-types}
\begin{tabular}{p{0.22\linewidth} p{0.68\linewidth}}
\toprule
\textbf{Type} & \textbf{Description} \\
\midrule
\texttt{random\_mono} &
Randomly placed spheres with similar radii. \\
\texttt{random\_poly} &
Randomly placed spheres with heterogeneous radii. \\
\texttt{channel} &
Spheres arranged so that a connected, relatively open pore pathway forms along one direction. \\
\texttt{layered} &
Spheres arranged with spatial density varying by layers along one direction, without necessarily forming a continuous open channel. \\
\texttt{bottleneck} &
Structures containing narrow throats or bottleneck regions in the pore space. \\
\texttt{clustered} &
Spheres spatially clustered in selected regions of the domain. \\
\bottomrule
\end{tabular}
\end{table}

The geometry is voxelized on a regular grid of resolution $N^3$, with
$N\in\{16,32\}$. The voxelized pore domain $P_h$ is then discretized by a
structured $P_1$ tetrahedral finite-element mesh obtained by subdividing each
pore voxel into six tetrahedra. Thus, if $N_{\mathrm{pore}}$ is the number of
pore voxels, then the number of tetrahedra is approximately $ N_{\mathrm{tet}} \approx 6N_{\mathrm{pore}}.$

The grid controls the voxelization resolution of the geometry, while the
diffusion equation is solved on the resulting tetrahedral finite-element mesh.
In all simulations, the release curve is computed over a prescribed finite time
interval $[0,t_{\mathrm{end}}]$, where $t_{\mathrm{end}}$ is fixed in advance.

For each simulated release curve $Q(t)$, we use three curve-derived summaries:
\begin{enumerate}[label=(\roman*)]
    \item $Q_{\mathrm{early}} = Q(t_{20\%})$,
    \item $Q_{\mathrm{mid}} = Q(t_{50\%})$,
    \item $Q_{\mathrm{final}} = Q(t_{\mathrm{end}})$.
\end{enumerate}
Here $t_{20\%}$ and $t_{50\%}$ denote 20\% and 50\% of the prescribed
simulation time interval, respectively. 

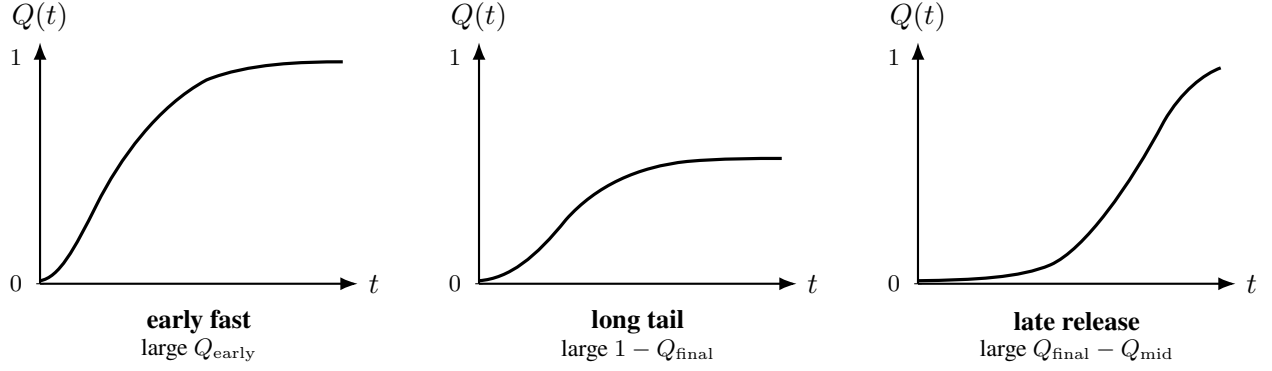
\begin{figure}[t]
\centering
\begin{tikzpicture}[
    >=Latex,
    axis/.style={->, thick},
    curve/.style={very thick},
    ticklabel/.style={font=\footnotesize},
    paneltitle/.style={font=\small\bfseries, align=center},
    note/.style={font=\footnotesize, align=center}
]

\newcommand{\releaseaxispanel}[2]{
    \draw[axis] (0,0) -- (4.2,0) node[right] {$t$};
    \draw[axis] (0,0) -- (0,3.2) node[above] {$Q(t)$};

    \draw (-0.05,0) -- (0.05,0);
    \node[ticklabel, anchor=east] at (-0.10,0) {$0$};

    \draw (-0.05,3) -- (0.05,3);
    \node[ticklabel, anchor=east] at (-0.10,3) {$1$};

    \node[paneltitle] at (2.1,-0.50) {#1};
    \node[note] at (2.1,-0.88) {#2};
}

\begin{scope}[shift={(-5.8,0)}]
    \releaseaxispanel{early fast}{large $Q_{\mathrm{early}}$}

    \draw[curve]
        (0,0.04)
        .. controls (0.25,0.08) and (0.45,0.45) .. (0.75,1.05)
        .. controls (1.05,1.65) and (1.55,2.35) .. (2.20,2.70)
        .. controls (2.75,2.92) and (3.45,2.94) .. (4.0,2.94);
\end{scope}

\begin{scope}[shift={(0,0)}]
    \releaseaxispanel{long tail}{large $1-Q_{\mathrm{final}}$}

    \draw[curve]
        (0,0.04)
        .. controls (0.40,0.07) and (0.75,0.35) .. (1.15,0.85)
        .. controls (1.55,1.30) and (2.10,1.55) .. (2.75,1.62)
        .. controls (3.20,1.66) and (3.65,1.66) .. (4.0,1.66);

\end{scope}

\begin{scope}[shift={(5.8,0)}]
    \releaseaxispanel{late release}
    {large $Q_{\mathrm{final}}-Q_{\mathrm{mid}}$}

    \draw[curve]
        (0,0.04)
        .. controls (0.75,0.05) and (1.35,0.08) .. (1.75,0.25)
        .. controls (2.20,0.45) and (2.80,1.30) .. (3.25,2.15)
        .. controls (3.48,2.55) and (3.78,2.78) .. (4.0,2.86);

\end{scope}

\end{tikzpicture}

\caption{
Schematic illustration of the three release regimes used in this study.
Early-fast samples are characterized by large early release
$Q_{\mathrm{early}}$, long-tail samples by a large final unreleased fraction
$1-Q_{\mathrm{final}}$, and late-release samples by a large late-time increase
$Q_{\mathrm{final}}-Q_{\mathrm{mid}}$.
}
\label{fig:release-regimes}
\end{figure}

For each porosity condition, we generated 720 candidate samples. The
\texttt{early\_fast} regime was first defined as the top 25\% of samples ranked
by $Q_{\mathrm{early}}$, yielding 180 samples. After removing these samples, the
\texttt{long\_tail} regime was defined among the remaining 540 samples as the
180 samples with the largest values of $1-Q_{\mathrm{final}}$. After removing
both the \texttt{early\_fast} and \texttt{long\_tail} samples, the
\texttt{late\_release} regime was defined among the remaining 360 samples as the
180 samples with the largest values of $Q_{\mathrm{final}}-Q_{\mathrm{mid}}$.
Thus, the three regimes were selected sequentially and each contained 180
samples. The remaining 180 samples were left unassigned and were not used in
the regime-wise statistical tests or classification analyses.
Figure~\ref{fig:release-regimes} gives a schematic illustration of these
regimes.

\subsection{Porosity-Controlled Association Between Release-Curve Regimes and Topological Summaries}
\label{subsec:porosity-controlled-curve-type-pd}

We next ask whether the association between release behavior and topological
summaries persists after controlling for total porosity. Using the
release-regime definitions introduced in
Figure~\ref{fig:release-regimes}, we performed the analysis separately within
the fixed target-porosity subsets $40\%$, $60\%$, and $80\%$. Thus, the
comparison is made only among samples with the same prescribed porosity level,
rather than across the full dataset.

For each fixed porosity level, we used the three disjoint release-regime groups
defined in Section~\ref{subsec:experimental-setting}. Each group contains 180
samples, so each fixed porosity comparison uses $n=540$ samples and $k=3$
groups.

For each sample and for each homology group $H_0$, $H_1$, and $H_2$, we
computed six persistence-diagram summaries: count, finite count, persistence sum, mean persistence, maximum persistence, and mean birth scale. This gives $3\times 6=18$ topological summaries (see \Cref{tab:persistence-summary-features}). 

\begin{table}[t]
\centering
\small
\renewcommand{\arraystretch}{1.55}
\caption{Persistence-diagram summary features used in this study. For each
homology dimension $q$, let $\mathrm{Dgm}_q$ denote the persistence diagram and
let
$\mathrm{Dgm}_q^{\mathrm{fin}}
=\{(b,d)\in \mathrm{Dgm}_q : d<\infty\}$
denote the set of features with finite death values. When
$\mathrm{Dgm}_q^{\mathrm{fin}}$ is empty, the mean and maximum summaries are set
to zero in the numerical implementation.}
\label{tab:persistence-summary-features}
\begin{tabular}{p{0.22\textwidth} p{0.34\textwidth} p{0.36\textwidth}}
\toprule
Feature & Formula & Description \\
\midrule

\texttt{count}
&
$|\mathrm{Dgm}_q|$
&
Total number of persistence features in homology dimension $q$. \\
\addlinespace[0.35em]

\texttt{finite\_count}
&
$|\mathrm{Dgm}_q^{\mathrm{fin}}|$
&
Number of persistence features with finite death values. \\
\addlinespace[0.35em]

\texttt{persistence\_sum}
&
$\displaystyle
\sum_{(b,d)\in \mathrm{Dgm}_q^{\mathrm{fin}}}(d-b)$
&
Sum of persistence lifetimes over features with finite death values. \\
\addlinespace[0.35em]

\texttt{persistence\_mean}
&
$\displaystyle
\frac{1}{|\mathrm{Dgm}_q^{\mathrm{fin}}|}
\sum_{(b,d)\in \mathrm{Dgm}_q^{\mathrm{fin}}}(d-b)$
&
Mean persistence lifetime over features with finite death values. \\
\addlinespace[0.35em]

\texttt{persistence\_max}
&
$\displaystyle
\max_{(b,d)\in \mathrm{Dgm}_q^{\mathrm{fin}}}(d-b)$
&
Maximum persistence lifetime among features with finite death values. \\
\addlinespace[0.35em]

\texttt{birth\_mean}
&
$\displaystyle
\frac{1}{|\mathrm{Dgm}_q^{\mathrm{fin}}|}
\sum_{(b,d)\in \mathrm{Dgm}_q^{\mathrm{fin}}} b$
&
Mean birth filtration value over features with finite death values. \\

\bottomrule
\end{tabular}
\end{table}

\begin{figure}[ht!]
    \centering

    \includegraphics[width=\textwidth]{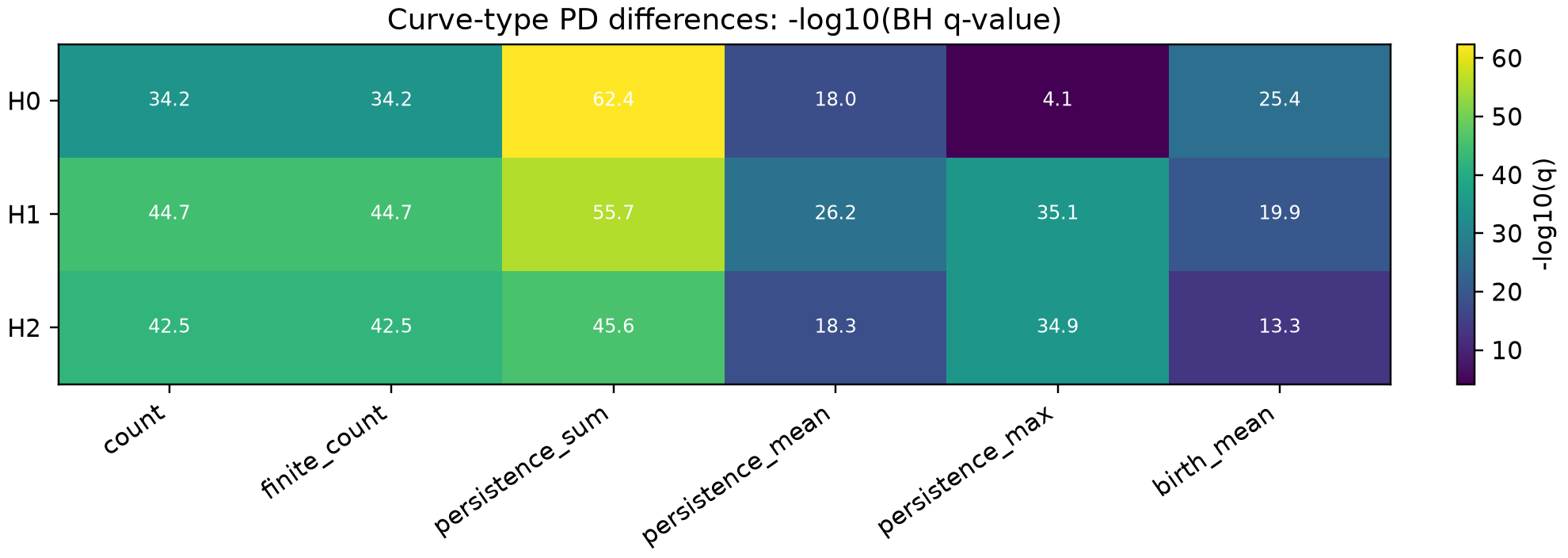}

    \vspace{0.5em}

    \includegraphics[width=\textwidth]{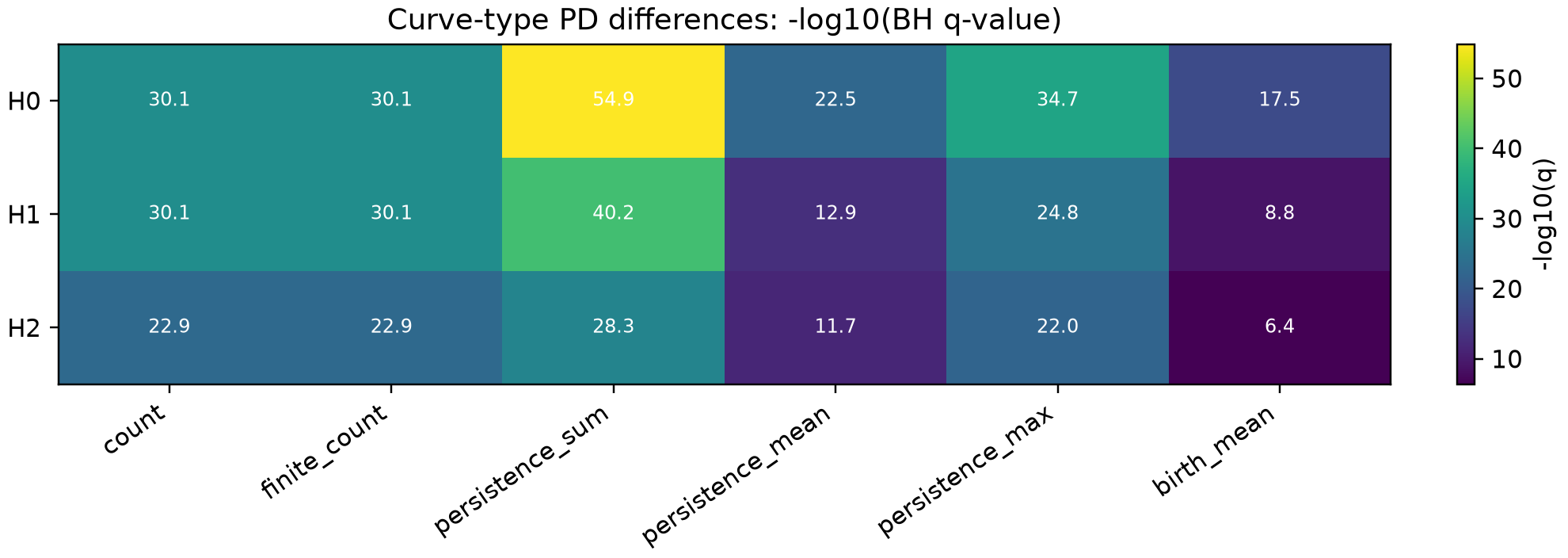}

    \vspace{0.5em}

    \includegraphics[width=\textwidth]{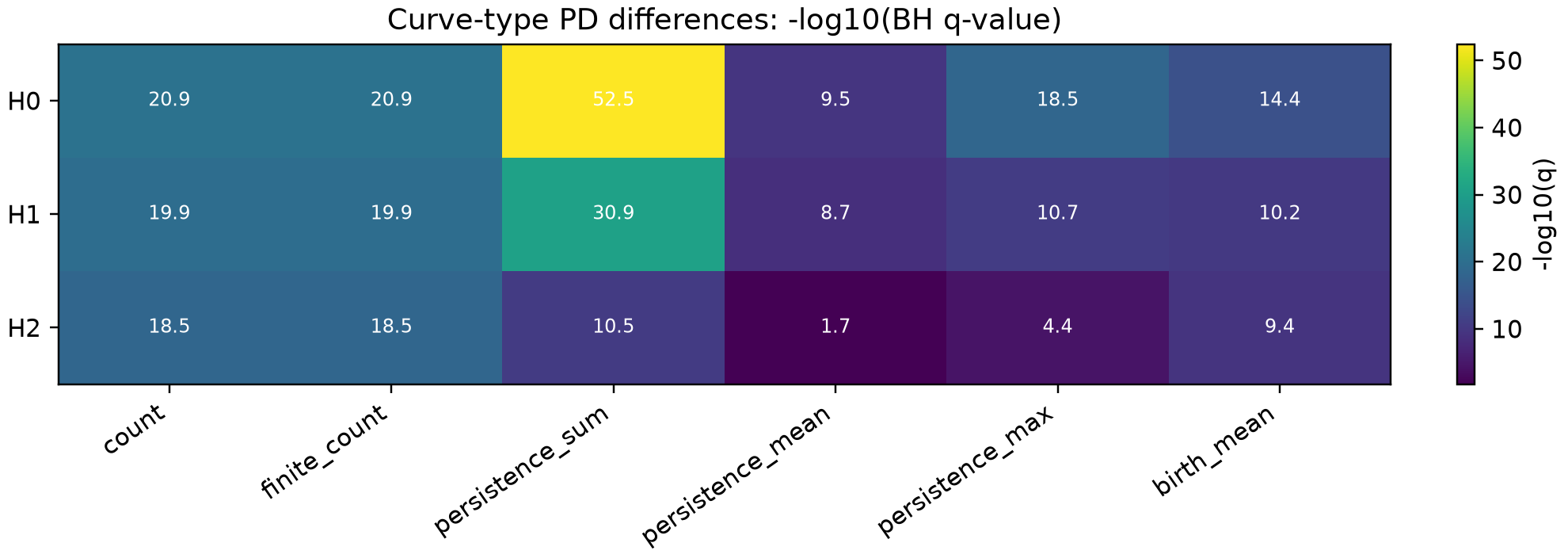}

    \caption{Porosity-controlled association between release-curve regimes and
    persistence-diagram summaries. From top to bottom, the panels correspond to
    target porosity levels of $40\%$, $60\%$, and $80\%$. Each entry shows
    $-\log_{10}(q)$, where $q$ is the Benjamini--Hochberg corrected $q$-value
    from a Kruskal--Wallis test across the three release-curve regimes.}
    \label{fig:porosity-controlled-curve-type-pd}
\end{figure}

For each fixed porosity level and each persistence summary, we used the
Kruskal--Wallis test~\cite{kruskal1952use} to test whether the three
release-curve regimes have the same distribution. Since 18 persistence
summaries were tested within each porosity level, the resulting $p$-values were
corrected using the Benjamini--Hochberg procedure~\cite{benjamini1995controlling}.
The heatmaps in Figure~\ref{fig:porosity-controlled-curve-type-pd} report
$-\log_{10}(q)$, where $q$ is the corrected $q$-value.

We also report the Kruskal--Wallis eta-squared effect size based on the $H$ statistic,
\[
    \eta_H^2
    =
    \frac{H_{\mathrm{KW}}-k+1}{n-k},
\]
where $H_{\mathrm{KW}}$ is the Kruskal--Wallis statistic, $n$ is the number of
observations, and $k$ is the number of groups
\cite{tomczak2014effectsize,kassambara2023rstatix}. This effect size is useful
because very small corrected $q$-values can occur in large samples even when
the magnitude of separation is modest.

\begin{figure}
    \centering
    \includegraphics[width=\linewidth]{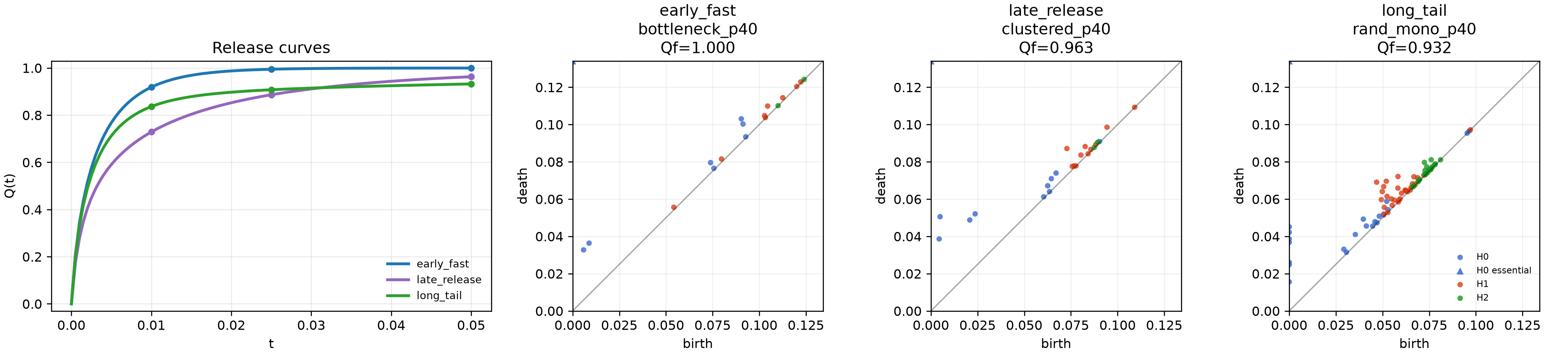}
    \caption{
    Representative release curves and weighted-alpha persistence diagrams for
    the three release regimes. The left panel compares the cumulative release
    curves $Q(t)$ for early-fast, late-release, and long-tail examples. The
    remaining panels show the corresponding persistence diagrams in dimensions
    $H_0$, $H_1$, and $H_2$. These examples provide visual intuition for the
    regime-wise topological differences; the statistical evidence is quantified
    in Figure~\ref{fig:porosity-controlled-curve-type-pd} and
    Table~\ref{tab:curve-type-kruskal-effect-sizes}.
    }
    \label{fig:release-regime-pd-examples}
\end{figure}

\begin{table}[t]
\centering
\caption{Kruskal--Wallis significance and effect size for topological summaries
at fixed porosity level. Each entry reports $-\log_{10}(q)$ followed by $\eta_H^2$.
S, M, and L denote small, medium, and large effect sizes, respectively.}
\label{tab:curve-type-kruskal-effect-sizes}
\scriptsize
\resizebox{0.72\textwidth}{!}{%
\begin{tabular}{l c c c}
\toprule
Persistence summary
& $40\%$ porosity
& $60\%$ porosity
& $80\%$ porosity \\
\midrule
$H_0$ count
& $34.2;\; 0.291$ (L)
& $30.1;\; 0.258$ (L)
& $20.9;\; 0.181$ (L) \\
$H_0$ finite count
& $34.2;\; 0.291$ (L)
& $30.1;\; 0.258$ (L)
& $20.9;\; 0.181$ (L) \\
$H_0$ pers. sum
& $62.4;\; 0.542$ (L)
& $54.9;\; 0.478$ (L)
& $52.5;\; 0.457$ (L) \\
$H_0$ pers. mean
& $18.0;\; 0.151$ (L)
& $22.5;\; 0.191$ (L)
& $9.5;\; 0.079$ (M) \\
$H_0$ pers. max
& $4.1;\; 0.032$ (S)
& $34.7;\; 0.301$ (L)
& $18.5;\; 0.159$ (L) \\
$H_0$ birth mean
& $25.4;\; 0.215$ (L)
& $17.5;\; 0.148$ (L)
& $14.4;\; 0.122$ (M) \\
\midrule
$H_1$ count
& $44.7;\; 0.384$ (L)
& $30.1;\; 0.259$ (L)
& $19.9;\; 0.171$ (L) \\
$H_1$ finite count
& $44.7;\; 0.384$ (L)
& $30.1;\; 0.259$ (L)
& $19.9;\; 0.171$ (L) \\
$H_1$ pers. sum
& $55.7;\; 0.482$ (L)
& $40.2;\; 0.349$ (L)
& $30.9;\; 0.269$ (L) \\
$H_1$ pers. mean
& $26.2;\; 0.222$ (L)
& $12.9;\; 0.108$ (M)
& $8.7;\; 0.071$ (M) \\
$H_1$ pers. max
& $35.1;\; 0.300$ (L)
& $24.8;\; 0.211$ (L)
& $10.7;\; 0.090$ (M) \\
$H_1$ birth mean
& $19.9;\; 0.168$ (L)
& $8.8;\; 0.072$ (M)
& $10.2;\; 0.085$ (M) \\
\midrule
$H_2$ count
& $42.5;\; 0.364$ (L)
& $22.9;\; 0.194$ (L)
& $18.5;\; 0.157$ (L) \\
$H_2$ finite count
& $42.5;\; 0.364$ (L)
& $22.9;\; 0.194$ (L)
& $18.5;\; 0.157$ (L) \\
$H_2$ pers. sum
& $45.6;\; 0.394$ (L)
& $28.3;\; 0.242$ (L)
& $10.5;\; 0.088$ (M) \\
$H_2$ pers. mean
& $18.3;\; 0.154$ (L)
& $11.7;\; 0.097$ (M)
& $1.7;\; 0.011$ (S) \\
$H_2$ pers. max
& $34.9;\; 0.298$ (L)
& $22.0;\; 0.186$ (L)
& $4.4;\; 0.035$ (S) \\
$H_2$ birth mean
& $13.3;\; 0.110$ (M)
& $6.4;\; 0.051$ (S)
& $9.4;\; 0.077$ (M) \\
\bottomrule
\end{tabular}%
}
\end{table}

The results show that topological summaries remain strongly associated with
release regimes even after porosity stratification. The persistence sum is the
most stable signal. The $H_0$ persistence sum has $-\log_{10}(q)$ values
62.4, 54.9, and 52.5 at porosity levels $40\%$, $60\%$, and $80\%$,
respectively, with large effect sizes $\eta_H^2=0.542$, $0.478$, and
$0.457$. The corresponding $H_1$ persistence sum values are
$-\log_{10}(q)=55.7$, 40.2, and 30.9, again with large effect sizes
$\eta_H^2=0.482$, $0.349$, and $0.269$. The $H_2$ persistence sum is also
significant at all porosity levels, although its effect size decreases to a
medium level at $80\%$ porosity.

The strength of separation decreases with porosity. At $40\%$ porosity, 16 out
of 18 summaries have large effect sizes, whereas at $60\%$ and $80\%$ porosity
the corresponding numbers are 14 and 9. This trend is most visible in
higher-dimensional summaries. For example, the effect size of $H_2$ persistence sum decreases from $\eta_H^2=0.394$ at $40\%$ porosity to
$\eta_H^2=0.088$ at $80\%$ porosity, and the effect size of $H_2$ maximum
persistence decreases from $\eta_H^2=0.298$ to $\eta_H^2=0.035$. This is
consistent with the fact that high-porosity samples contain a sparser solid
phase, so loop- and cavity-level solid-offset features become simpler and less
discriminative.

\begin{table}[t]
\centering
\caption{Representative group means for automatically defined release-curve
regimes and selected topological summaries at fixed porosity level. The heatmaps in
Figure~\ref{fig:porosity-controlled-curve-type-pd} and the effect sizes in
Table~\ref{tab:curve-type-kruskal-effect-sizes} test whether the groups differ,
while this table shows the direction of the differences.}
\label{tab:curve-type-group-summary-key}
\scriptsize
\resizebox{\textwidth}{!}{%
\begin{tabular}{c l l c c c c}
\toprule
Porosity & Curve regime & Defining curve summary
& $H_0$ pers. sum & $H_1$ pers. sum & $H_2$ pers. sum & $H_1$ count \\
\midrule
$40\%$ & early\_fast
& $Q_{\mathrm{early}}=0.903$
& 0.137 & 0.0372 & 0.00447 & 15.36 \\
$40\%$ & late\_release
& $\mathrm{late\_gain}=0.0263$
& 0.151 & 0.0430 & 0.00498 & 15.93 \\
$40\%$ & long\_tail
& $\mathrm{unreleased\_final}=0.0399$
& 0.250 & 0.1059 & 0.01723 & 29.44 \\
\midrule
$60\%$ & early\_fast
& $Q_{\mathrm{early}}=0.734$
& 0.110 & 0.0384 & 0.00483 & 19.28 \\
$60\%$ & late\_release
& $\mathrm{late\_gain}=0.0799$
& 0.175 & 0.0662 & 0.00855 & 29.77 \\
$60\%$ & long\_tail
& $\mathrm{unreleased\_final}=0.0718$
& 0.211 & 0.0859 & 0.01312 & 34.26 \\
\midrule
$80\%$ & early\_fast
& $Q_{\mathrm{early}}=0.593$
& 0.152 & 0.0635 & 0.00672 & 17.95 \\
$80\%$ & late\_release
& $\mathrm{late\_gain}=0.1282$
& 0.230 & 0.1022 & 0.01137 & 31.65 \\
$80\%$ & long\_tail
& $\mathrm{unreleased\_final}=0.1101$
& 0.248 & 0.1146 & 0.01265 & 28.76 \\
\bottomrule
\end{tabular}%
}
\end{table}

The group means in Table~\ref{tab:curve-type-group-summary-key} show the
direction of the differences. Across all three porosity levels, early-fast
samples have the smallest persistence sum and the smallest $H_1$ counts,
whereas long-tail samples have the largest or nearly largest values. For
example, at $40\%$ porosity, the $H_0$, $H_1$, and $H_2$ persistence sum
values increase from 0.137, 0.0372, and 0.00447 in the early-fast group to
0.250, 0.1059, and 0.01723 in the long-tail group. At $60\%$ and $80\%$
porosity, the same pattern remains visible, with the late-release group usually
lying between early-fast and long-tail behavior.

Overall, the porosity-controlled analysis shows that the relationship between
persistent homology and release-curve shape is not explained solely by total
porosity. Even within fixed porosity levels, samples with different release
regimes exhibit statistically distinct topological summaries. The most robust
pattern is that early-fast release is associated with smaller persistence
counts and smaller persistence sum, whereas long-tail release is associated
with larger persistence counts and larger persistence sum.

In physical terms, these results indicate that, even at the same prescribed
porosity, more complex solid-phase topology is associated with slower or more
persistent release behavior. Larger persistence counts and larger persistence sum reflect a more intricate multiscale organization of connected
components, loop-like structures, and cavity-like structures in the solid-offset
filtration. These topological features are most pronounced in long-tail samples
and weakest in early-fast samples, suggesting that release behavior depends not
only on the amount of pore space but also on how the solid phase is organized
across scales.

The release-curve regimes are defined from $Q(t)$ itself, and the test
asks whether topological summaries differ across these curve-derived groups.
Therefore, the result shows that topological summaries are associated with
release behavior after porosity stratification, but it does not by itself prove
that these summaries causally determine the release curve.

\subsection{Topology-Based Classification of Release Regimes}
\label{subsec:regime-classification}

The regime-wise association analysis shows that topological summaries vary
systematically across release regimes. We next ask whether these topological
summaries are sufficient to classify release behavior from geometry alone. In
this experiment, each sample is represented solely by the 18
persistent homology summary features introduced in
\Cref{subsec:porosity-controlled-curve-type-pd}.

We consider a three-class classification task over the three release regimes:
\texttt{early\_fast}, \texttt{late\_release}, and \texttt{long\_tail}.

For each grid--porosity setting, we train a multinomial logistic regression
classifier on the 18 persistent homology summaries
\cite{mcfadden1974conditional,agresti2013categorical}. We use this linear
multinomial model, rather than a more flexible nonlinear classifier, to keep the
classification analysis interpretable: the fitted coefficients can be directly
related to the contribution of each persistence summary to the separation of
release regimes. Classification experiments are repeated over five stratified
random train--test splits. In each split, 75\% of the samples are used for
training and 25\% are held out for testing. Since the three classes are balanced
and the splits are stratified, accuracy and balanced accuracy coincide.

\begin{table}[t]
\centering
\caption{
Topology-based classification of three release regimes using only
18 persistent homology summary features. Each entry reports the test-set mean
$\pm$ standard deviation over five random train--test splits.
}
\label{tab:regime-classification}
\begin{tabular}{c c c c c}
\toprule
\textbf{Grid} &
\textbf{Porosity} &
\textbf{Accuracy} &
\textbf{Balanced accuracy} &
\textbf{Macro F1} \\
\midrule
$16^3$ & $40\%$ & $0.759 \pm 0.023$ & $0.759 \pm 0.023$ & $0.752 \pm 0.023$ \\
$16^3$ & $60\%$ & $0.683 \pm 0.026$ & $0.683 \pm 0.026$ & $0.683 \pm 0.024$ \\
$16^3$ & $80\%$ & $0.641 \pm 0.028$ & $0.641 \pm 0.028$ & $0.639 \pm 0.026$ \\
$32^3$ & $40\%$ & $0.738 \pm 0.023$ & $0.738 \pm 0.023$ & $0.739 \pm 0.019$ \\
$32^3$ & $60\%$ & $0.653 \pm 0.018$ & $0.653 \pm 0.018$ & $0.653 \pm 0.017$ \\
$32^3$ & $80\%$ & $0.668 \pm 0.031$ & $0.668 \pm 0.031$ & $0.667 \pm 0.032$ \\
\bottomrule
\end{tabular}
\end{table}

The classification results show that persistent homology summaries contain
substantial information about the terminal release regime. Across all evaluated
grid--porosity settings, the test accuracy ranges from $0.641$ to $0.759$,
which is well above the chance level of $1/3$ for a balanced three-class task.
The corresponding macro F1 scores range from $0.639$ to $0.752$, indicating
that the classifier does not achieve its performance by favoring only one
regime.

The strongest classification performance is observed at $40\%$ porosity. At
this porosity level, the test accuracy is $0.759 \pm 0.023$ for the $16^3$ grid
and $0.738 \pm 0.023$ for the $32^3$ grid. Performance decreases at higher
porosity levels, with accuracies between approximately $0.64$ and $0.68$ for
$60\%$ and $80\%$ porosity. This trend is consistent with the
porosity-controlled association analysis: when the solid phase becomes sparse,
higher-dimensional topological features become less discriminative, and release
regimes are harder to separate using topology alone.
These results suggest that
persistent homology summaries capture release-regime information directly from
the multiscale organization of the solid phase.

\subsection{Computational Timing}
\label{subsec:computational-timing}

We also measured the average runtime required for the main stages of
the simulation and feature-extraction pipeline. All timings are reported in
seconds. Runtime measurements were performed on a Windows 11 workstation equipped with
an Intel Core Ultra 7 255H CPU, 16 logical processors, and 31.4 GB RAM.
Experiments used Python~3.12.13 with NumPy~2.5.0, SciPy~1.18.0, and
GUDHI~3.12.0 \cite{harris2020array,virtanen2020scipy,gudhi:urm}.

For each grid--porosity setting, timings were averaged over 100 samples. The
diffusion solve computes the release curve $Q(t)$ at 101 time points using the
tetrahedral finite-element scheme. The release curve is therefore computed on
the voxel-derived tetrahedral mesh. In contrast, persistent homology is computed
directly from the weighted sphere representation used to define the solid-offset
filtration. The topological feature time is defined as the sum of the
weighted-alpha persistent homology computation and the extraction of the 18
scalar topological summaries.
Table~\ref{tab:timing-summary} reports the mean wall-clock time per sample in
seconds, with standard deviations in parentheses.

\begin{table}[t]
\centering
\caption{Average wall-clock computation time per sample for the main pipeline
components. Each entry reports mean time in seconds, with standard deviation in
parentheses, over 100 samples. Runtime measurements were performed on the
workstation described in the text. Persistence images are excluded. The final
column reports the ratio between the Tet-FEM release-curve solve and scalar PH
feature extraction.}
\label{tab:timing-summary}
\scriptsize
\resizebox{\textwidth}{!}{%
\begin{tabular}{c c c c c c c c}
\toprule
Grid
& Porosity
& Geometry (s)
& Voxelization (s)
& Tet-FEM $Q(t)$ (s)
& Weighted-alpha PH (s)
& PH summaries (s)
& Tet-FEM / PH features \\
\midrule
$N=16$ & $40\%$
& $0.0373\;(0.0114)$
& $0.000004\;(0.000001)$
& $0.1180\;(0.0084)$
& $0.0177\;(0.0051)$
& $0.000092\;(0.000049)$
& $6.6\times$ \\

$N=16$ & $60\%$
& $0.0200\;(0.0067)$
& $0.000003\;(0.000001)$
& $0.2010\;(0.0153)$
& $0.0084\;(0.0035)$
& $0.000071\;(0.000019)$
& $23.8\times$ \\

$N=16$ & $80\%$
& $0.0123\;(0.0049)$
& $0.000003\;(0.000001)$
& $0.3007\;(0.0253)$
& $0.0046\;(0.0024)$
& $0.000076\;(0.000046)$
& $64.7\times$ \\

\midrule

$N=32$ & $40\%$
& $0.1298\;(0.0339)$
& $0.000019\;(0.000007)$
& $1.4362\;(0.0818)$
& $0.0175\;(0.0048)$
& $0.000092\;(0.000042)$
& $81.7\times$ \\

$N=32$ & $60\%$
& $0.0723\;(0.0213)$
& $0.000018\;(0.000007)$
& $2.2786\;(0.0966)$
& $0.0084\;(0.0034)$
& $0.000081\;(0.000016)$
& $269.3\times$ \\

$N=32$ & $80\%$
& $0.0442\;(0.0144)$
& $0.000018\;(0.000007)$
& $3.0555\;(0.0672)$
& $0.0048\;(0.0025)$
& $0.000074\;(0.000015)$
& $631.0\times$ \\
\bottomrule
\end{tabular}%
}
\end{table}

The timing results show that the finite-element release simulation is the
dominant computational cost, especially at the higher grid resolution. For
$N=16$, the Tet-FEM solve requires between 0.118 and 0.301 seconds per sample,
whereas scalar PH feature extraction requires between approximately 0.0047 and
0.0178 seconds per sample. For $N=32$, the difference becomes more pronounced:
the Tet-FEM solve requires between 1.44 and 3.06 seconds per sample, while
scalar PH feature extraction remains below 0.018 seconds per sample.

Consequently, the Tet-FEM solve is about $6.6$ to $64.7$ times slower than
scalar PH feature extraction on the $16^3$ grid, and about $81.7$ to $631.0$
times slower on the $32^3$ grid. This supports the interpretation of persistent
homology as a lightweight geometry-based screening descriptor. The topological
summaries are not intended to replace the release simulation, but they can be
computed at much lower cost and may help prioritize candidate geometries before
more detailed PDE-based analysis.

\section{Conclusion}

We used persistent homology to quantify the multiscale topological and geometric
organization of porous media, including solid connectivity and the formation of
loop-like and cavity-like structures across spatial scales. Our results show
that release regimes have distinct topological signatures even after controlling
for porosity. In particular, more complex solid-phase topology is associated
with long-tail release behavior.

The classification and timing experiments further support the use of persistent
homology as a lightweight and interpretable geometry-based screening descriptor.
Persistent homology-based features can distinguish release-curve regimes using
a simple classification model, and their extraction requires substantially less
computational time than FEM diffusion simulations. Thus, persistent homology
should be viewed as a complementary tool for prioritizing candidate porous
structures before more expensive simulation-based analysis.

Future work will extend this framework beyond synthetic spherical-obstacle
media to experimentally measured porous microstructures. In particular, an
important direction is to apply the proposed method to pharmaceutical porous
systems, such as sustained-release tablets, where pore architecture strongly
affects drug release profiles. Combining persistent homology descriptors with
micro-CT data, formulation variables, and experimentally measured dissolution
curves may provide an interpretable way to analyze and screen drug-delivery
materials with desired release behavior.

\bibliographystyle{unsrtnat}
\bibliography{references}

\end{document}